\documentclass[12pt]{iopart}

\usepackage{graphicx}
\begin{document}

\title[Excluded volume and molecular field in the Lennard-Jones fluid]
{Excluded volume and molecular field in the Lennard-Jones fluid: a modified first-order perturbation theory}

\author{A. Trokhymchuk$^{1,2}$\footnote{Corresponding author}, V. Hordiichuk$^{1}$, R. Melnyk$^{1}$, I. Nezbeda$^{3,4}$}

\address{
$^1$Yukhnovskii Institute for Condensed Matter Physics of the National Academy of Sciences of
Ukraine, 79011 Lviv, Ukraine\\
$^2$Faculty of Chemistry and Chemical Technology, University of Ljubljana, 1000 Ljubljana, Slovenia\\
$^3$Faculty of Science, J. E. Purkinje University,  400 96 \'{U}st\'{\i} nad Labem, Czech Republic \\
$^4$E. H\'ala Laboratory of Thermodynamics, Institute of Chemical Process Fundamentals of  Academy of Sciences,
165 02 Prague 6, Czech Republic
}

\ead{adt@icmp.lviv.ua}
\vspace{10pt}
\begin{indented}
\item[]April 2026
\end{indented}

\begin{abstract}
The equation of state and, more generally, the thermodynamics of the Lennard-Jones fluid have long served as a benchmark problem in the statistical theory of fluids. Among available theoretical approaches, first-order perturbation theory occupies a special position: only at this level does the correction to the Helmholtz free energy admit an exact statistical-mechanical expression. In this work, we present a systematic, simulation-based assessment of a non-classical first-order perturbation theory in which the reference system incorporates the entire short-range part of the interaction, while the perturbation is confined to the remaining long-range tail.

We show that this range-based decomposition transforms the perturbation contribution into a small, smoothly varying, near-mean-field quantity over a broad supercritical thermodynamic domain. When its density and temperature derivatives are consistently retained, the resulting equation of state reproduces high-accuracy reference data with excellent fidelity. The results demonstrate that the success of first-order perturbation theory is governed primarily by the physical content of the reference system and by the consistent treatment of its state dependence, rather than by the formal truncation order of the expansion.

\end{abstract}

%
\vspace{2pc}

\newpage

\section{\label{I}Introduction}

The Lennard-Jones (LJ) fluid occupies a unique position in the statistical mechanics of fluids. As the simplest model incorporating both short-range repulsion and long-range attraction, it has served for decades as a paradigmatic system for testing theoretical approaches to fluid thermodynamics, phase behavior, and equations of state. Owing to its simplicity and universality, the LJ model has been investigated extensively by computer simulations \cite{KolafaNezbeda1994,StephanFPE2020}, yielding highly accurate reference data over broad ranges of temperature and density. At the same time, it remains a central benchmark for assessing the validity, limitations, and internal consistency of theoretical descriptions of real fluids \cite{BH1976,WCA1980,HM1986,HM2013}.

Among the available theoretical approaches, perturbation theory plays a particularly prominent role \cite{BH1976,WCA1980,HM1986,HM2013}. In its general formulation, perturbation theory expresses the Helmholtz free energy of a fluid as an expansion around a suitably chosen reference system, with correction terms accounting for the remaining part of the intermolecular interactions. This framework underlies many classical and modern equations of state, including van der Waals-type theories \cite{vdwThesis}, Barker-Henderson (BH) theory \cite{BH1967}, Weeks-Chandler-Andersen (WCA) theory \cite{WCA1971}, as well as numerous semi-empirical and molecular-based models. Despite the development of higher-order perturbative schemes and sophisticated analytical treatments, a central and unresolved issue remains the choice of the reference system.

The special importance of first-order perturbation theory deserves emphasis, since only at this level does the perturbation contribution to the Helmholtz free energy admit an exact statistical-mechanical expression in terms of the reference-system structure \cite{BH1976,WCA1980,HM1986,HM2013,Zwanzig1954}. As a consequence, a successful first-order theory offers the prospect of a simple, transparent, and analytically tractable description of thermodynamic properties, without resorting to extensive parameterization or high-order corrections. Conversely, the failure of first-order perturbation theory signals a fundamental inadequacy in the underlying choice of the reference system rather than a mere quantitative deficiency.

In the classical approach (CA) to perturbation theory of the Lennard-Jones fluid, the reference system is defined by the purely repulsive part of the intermolecular potential, while the attractive interactions are treated as a perturbation \cite{Zwanzig1954}. In practical implementations such as the Barker-Henderson \cite{BH1967,BH1976} and Weeks-Chandler-Andersen \cite{WCA1971,WCA1980} theories, the soft repulsion of the Lennard-Jones potential is further mapped onto a hard-sphere reference with effective temperature dependent hard-core diameter. As a consequence, the properties of a reference system depend primarily on density. This construction leads naturally to a mean-field-type treatment of attractions. While, in principle, the first-order perturbation contribution in the classical splitting also acquires density and temperature dependence through the reference structure, retaining this dependence is not feasible in practice. The reason is that the perturbation interaction overlaps strongly with the region of pronounced structural correlations, and the use of the full reference radial distribution function would lead to large, strongly state-dependent corrections and poor convergence of the first-order expansion. As a result, classical perturbation theories are practically forced to adopt a mean-field approximation, a necessity that has been explicitly demonstrated by computer-based assessments of CA-type perturbation theories \cite{WestenGross2017}.

Motivated by these considerations, a non-classical approach (NCA) to the splitting of the intermolecular potential has been proposed \cite{MelnykFPE2009,MelnykJSF2010}. In this framework, the reference system is defined by the entire short-range part of the full potential, including not only the repulsive core but also the attractive well, while the perturbation is confined to the remaining long-range, weakly attractive tail. This choice is physically motivated by the observation that the dominant structural and thermodynamic features of dense fluids are largely determined by interactions within the first coordination shell (see also Ref. \cite{Dyre2011}). 
The NCA splitting introduces two fundamental and closely related novelties. 
First, the reference system incorporates not only excluded-volume effects but also the attractive interaction up to and including the potential minimum of the parent fluid. As a result, it captures the dominant short-range physics governing local structure, rather than relying on a hard-sphere representation of soft repulsion. This construction renders the reference system explicitly dependent on both density and temperature in a physically controlled manner.
Second, by confining the perturbation interaction to the remaining long-range tail beyond the first coordination shell, the magnitude of the perturbation term is substantially reduced. This separation by interaction range -- rather than by interaction strength -- allows the first-order perturbation contribution to remain weak and slowly varying, thereby restoring the viability of a consistent first-order perturbation treatment. 
The physical motivation and general implications of this range-based decomposition strategy were discussed in detail for hard-core and continuous model fluids in Ref.~\cite{MelnykJSF2010}, where it was shown that incorporating short-range attractive interactions into the reference system restores the viability of first-order perturbation theory.

An immediate and important consequence of a density- and temperature-dependent reference system is the appearance of an additional contribution to thermodynamic quantities derived from the Helmholtz free energy \cite{Zwanzig1954,MelnykFPE2009,MelnykJSF2010}. This extra term arises from derivatives of the perturbation contribution with respect to thermodynamic variables and has no counterpart in classical perturbation theories formulated at the mean-field level.

In classical first-order perturbation approaches based on the Barker-Henderson and Weeks-Chandler-Andersen splittings, the perturbation contribution is treated as a molecular-mean-field term and is therefore assumed to be independent of thermodynamic state variables. This restriction is not optional but is imposed by the strong overlap of the perturbation interaction with the region controlling local structure. In contrast, within the present non-classical framework, the perturbation contribution acquires an intrinsic state dependence, and a consistent first-order treatment requires that the resulting derivative terms be retained explicitly.

The NCA-based first-order perturbation theory has previously been applied to model hard-core based fluids with attractive Yukawa and Sutherland potentials \cite{MelnykFPE2009,MelnykJSF2010}. These studies demonstrated that, even when the perturbation term was evaluated within the mean-field approximation, the theory yields excellent agreement with simulation data over wide thermodynamic ranges, with noticeable deviations appearing primarily at high densities and extreme temperature conditions. These results provided strong evidence for the physical soundness of the NCA framework and pointed to the potential importance of the extra term in dense-fluid regimes.

The purpose of the present work is to provide a systematic and simulation-based assessment of the NCA-based first-order perturbation theory for the Lennard-Jones fluid. In contrast to previous applications relying partly on analytical approximations, we evaluated both the reference-system contribution and the first-order perturbation term directly by computer simulations. This strategy eliminates uncertainties associated with approximate theoretical descriptions and allows for a clean and unbiased test of the NCA framework itself. Particular attention is paid to the density and temperature dependence of the perturbation contribution and to the role of the extra thermodynamic term arising from its derivatives.

The paper is organized as follows. In Section 2 we briefly present: (1) the general formulation of perturbation theory for the Lennard-Jones fluid; (2) the classical and non-classical approaches to splitting the pair interaction potential into reference and perturbation parts; and (3) the first-order perturbation approximation, consisting of contributions from the reference system and from the perturbation. Section 3 is dedicated to results and discussions. The purpose of this section is to report on the assessment of the first order perturbation theory being applied to supercritical properties of the Lennard-Jones fluid. To avoid an impact of the approximate theoretical description, we evaluated all results for both contributions involving computer simulations. Section 4 summarizes the main conclusions and outlines implications for the development of molecular-based equations of state.

\section{\label{II} Statistical mechanical considerations}

\subsection{Perturbation theory formulation}
In the present study, we consider the fluid system with a pairwise additive Lennard-Jones (LJ) intermolecular potential
\begin{equation}
u(r) = 
4\epsilon \left [\left (\frac{\sigma_{\rm LJ}}{r}\right )^{12} - 
\left (\frac{\sigma_{\rm LJ}}{r}\right )^{6}\right ]\,,
\label{uLJ}
\end{equation}
where $r = |\mathbf{r}_i - \mathbf{r}_j|$ is the radial distance, $\,\epsilon\,$ is the depth of the potential well and $\,\sigma_{\rm LJ}\,$ is the characteristic Lennard-Jones diameter of the molecules. In simulations, we used $\sigma_{\rm LJ} = 3.405$~\AA{}, $\varepsilon=119.8$~K.

The application of perturbation theory to the thermodynamic description of this system is based on the assumption that intermolecular potential energy $\,U (\mathbf{r}_1,\mathbf{r}_2,\ldots,\mathbf{r}_N) = \sum_i^N\sum_{j>i}^N u(\mathbf{r}_i,\mathbf{r}_j)\,$  has to be split into two parts,
\begin{equation}  
U (\mathbf{r}_1,\mathbf{r}_2,\ldots,\mathbf{r}_N) = U_{\rm ref} (\mathbf{r}_1,\mathbf{r}_2,\ldots,\mathbf{r}_N) + \Delta U (\mathbf{r}_1,\mathbf{r}_2,\ldots,\mathbf{r}_N)
\label{ur}
\end{equation}
where $\,U_{\rm ref} (\mathbf{r}_1,\mathbf{r}_2,\ldots,\mathbf{r}_N) = \sum_i^N\sum_{j>i}^N u_{\rm ref}(\mathbf{r}_i,\mathbf{r}_j)\,$ and $\,\Delta U (\mathbf{r}_1,\mathbf{r}_2,\ldots,\mathbf{r}_N) = \sum_i^N\sum_{j>i}^N u_{\rm pert}(\mathbf{r}_i,\mathbf{r}_j)\,$ are referred to as the reference and perturbation parts, respectively, while $u_{\rm ref}(r)$ and $u_{\rm pert}(r)$ are the corresponding parts of the LJ potential (\ref{uLJ}),
\begin{equation}
u(r) = u_{\rm ref}(r) + u_{\rm pert}(r)\,.
\label{ur2}
\end{equation}
Then, for a small attractive perturbation, the system Helmholtz free energy $\,A\,$ can be expanded in powers of $\,u_{\rm pert}(r)/(kT)\,$, assuming the form
\begin{eqnarray}
\label{A_PT}
\frac{A}{kTN} &=& \frac{A_{\rm ref}}{kTN} 
+ \frac{\langle \Delta U \rangle_{\rm ref}}{kTN} - \frac{\langle [\Delta U - \langle\Delta U\rangle_{\rm ref}]^2 \rangle_{\rm ref}}{2!(kT)^2N} + \ldots \,,
\end{eqnarray}
where $\,\langle\ldots\rangle_{\rm ref}\,$ denotes a statistical average over a  canonical ensemble of the reference fluid at the same temperature $T$ and density $\rho=N/V$; all remaining notations are the following:
$\,N\,$ -- number of molecules, $\,V\,$ -- volume of the system, while 
 $\,k\,$ -- Boltzmann constant. 
The practical usefulness of expansion (\ref{A_PT}) is therefore controlled not by the formal smallness of $u_{\mathrm{pert}}$, but by the degree to which the reference system reproduces the local structure of the full fluid.

The first perturbation contribution in Eq.~(\ref{A_PT}) is the average interaction energy due to the pair perturbation interaction $u_{\rm pert}(r)$, while the second and higher-order terms represent fluctuations of this average and thus quantify the degree to which the reference system fails to capture the correlations induced by the full interaction. Therefore, the crucial step for the successful perturbation theory in general, and for the success of the first-order PT approach in particular, is the decomposition (\ref{ur2}). This decomposition is not unique and has traditionally been governed primarily by mathematical convenience rather than by considerations of convergence, whereas the physical issue of convergence of the perturbation expansion has received comparatively little attention. 

The necessary condition for the applicability of the theory is that the reference system reproduces the local structure of the considered fluid. 
Early simulation studies revealed that, for non-associating liquids, this condition is often satisfied by purely repulsive reference models. However, it is important to recall that this conclusion holds primarily for dense fluids, i.e., the liquid phase but not for low-density gas. 
At the same time, a serious drawback of this choice is that the thermodynamic behavior of hard body fluids is far from that of real fluids. One possibility to remove, at least partly, the above drawbacks is by incorporating in some way also the attractive interactions into the reference model which could extend the perturbation theory to the supercritical region, see e.g. \cite{MelnykFPE2009,MelnykJSF2010}. We therefore have  constructed a short-range LJ potential with an attractive part operating approximately within the first coordination shell~\cite{Hordiichuk2023}. 

Performed simulations confirmed the expectations: the model does reproduce the local structure not only of the dense liquid phase but also of the supercritical gaseous phase. Its thermodynamic properties are also evidently closer to those of the full LJ fluid. This implies that a theory based on such a reference should be considerably simpler, since the correction term need not compensate for deficiencies of the reference in describing the thermodynamic properties of the parent liquid. The correction term may attain thus a simple form, and such an equation should perform well over a wider range of thermodynamic conditions. These considerations naturally lead to a decomposition strategy in which the reference system is constructed to reproduce the short-range structure of the fluid, rather than being restricted to purely repulsive interactions, an idea that underlies the non-classical approach introduced below.

\subsection{Split of potential energy into reference and perturbation  parts}
\begin{figure}[htb]
	\centering
	\includegraphics[width=0.475\linewidth]{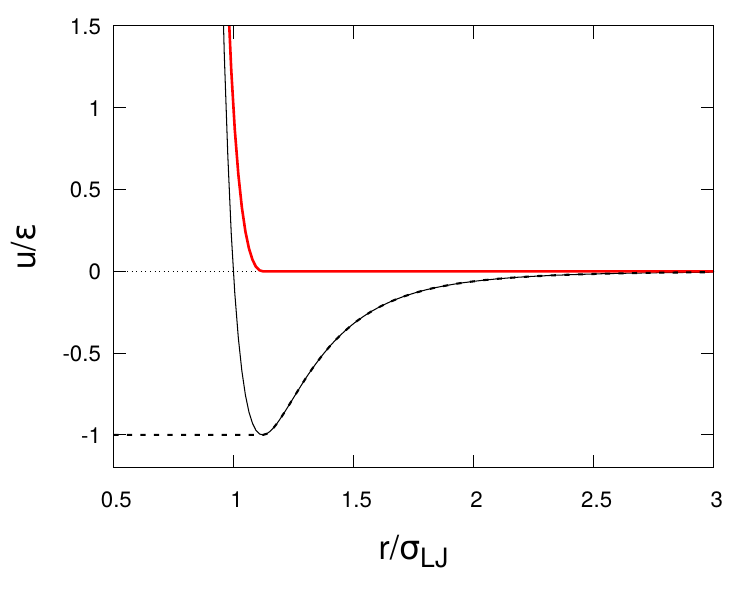}
    \includegraphics[width=0.475\linewidth]{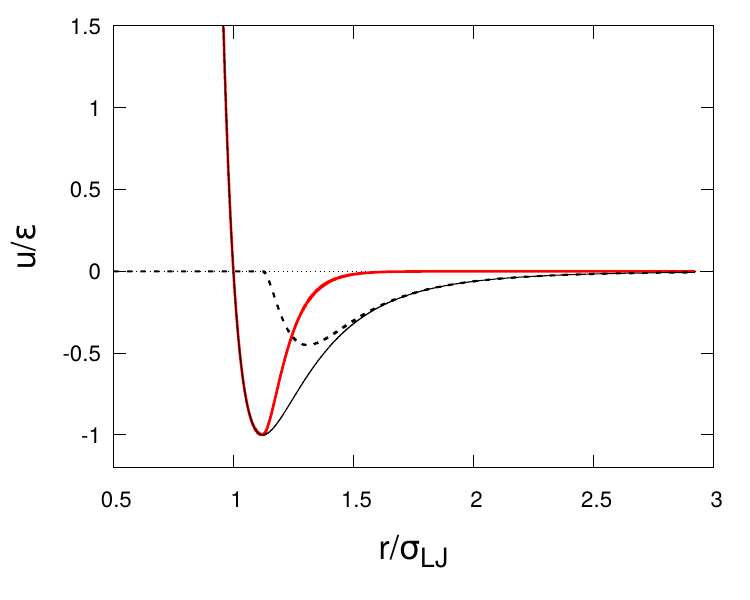}\\
	\caption{Lennard-Jones potential $\,u(r)\,$ (the thin solid line) and two choices of its decomposition. The left panel: The CA decomposition  suggested within the WCA method -- into the repulsive interaction (the thick solid line) and the attractive interaction (the thin dashed line). The right panel: The NCA decomposition suggested in Ref.~\cite{Hordiichuk2023} and used in present study -- into the short-range interaction $\,u_{\rm SR}(r)\,$ (the thick solid line) and the remaining attraction $\,\Delta u(r)\,$ (the thin dashed line). } \label{FigPOT}. 
\end{figure}

The classical approach (CA) to construct the reference and perturbation parts of the LJ potential (\ref{uLJ})  is due to Barker-Henderson (BH)~\cite{BH1967} and Weeks-Chandler-Andersen (WCA)~\cite{WCA1971} methods.
Despite minor technical variations, the central idea of CA is that the pair interaction energy $\,u(r)\,$ is divided into the repulsion energy, 
\begin{eqnarray}
\label{urep}
u_{\rm ref}(r) \equiv u_{\rm rep}(r) =\left\{
\begin{array}{ll}
u(r) + \epsilon, &  r\le r_{\rm m}  \\ \\
0, &  r > r_{\rm m}
\end{array}
\right.\,,  
\end{eqnarray}
and attraction energy $\,u_{\rm att}(r)\,$ contributions such that the full strength of LJ attraction energy becomes the perturbation, being just the difference 
\begin{equation}
u_{\rm pert}(r) \equiv u_{\rm att}(r) = u(r) - u_{\rm rep}(r)\,,
\label{uatt}
\end{equation}
where in Eq.~(\ref{urep}), the parameter $\,r_{\rm m}=2^{1/6}\sigma\,$ is the distance at which the LJ pair potential is at its minimum, while the constant shift ensures continuity of $u_{\mathrm{rep}}(r)$ at $r=r_{\mathrm m}\,$. In this construction, the entire attractive well contributes to the perturbation term, including the region of strongest correlations, where the reference radial distribution function deviates most strongly from unity.
Already from Fig.~\ref{FigPOT} (the left panel) one can see that with CA choice of the reference system, Eq.~(\ref{urep}), it will be very naive to expect the convergence of perturbation expansion (\ref{A_PT}) after the first perturbation correction. Indeed, recently reported the critical evaluation of CA-based perturbation theories by computer simulation \cite{WestenGross2017} found that none of the approaches studied leads to satisfactory results. 

The essential deficiency of the classical splitting is therefore not the magnitude of the perturbation energy but its spatial range, which overlaps strongly with the region controlling local structure. These shortcomings motivate a fundamentally different decomposition strategy, in which the reference system is defined by the interaction range rather than by repulsion alone. Therefore, in a recent study~\cite{Hordiichuk2023} the non-classical approach (NCA) to the reference interaction for the LJ potential $\,u(r)\,$  was reported. The non-classical approach to the reference interaction is based on extending the excluded-volume concept by incorporating short-range attractive interactions that act within the first coordination shell \cite{MelnykFPE2009,MelnykJSF2010}.  In these studies of the LJ-like hard-core attractive Yukawa and Sutherland fluids it was shown that using the hard-sphere repulsion plus short-range Yukawa reference system improves the first-order NCA perturbation theory to the level of the second-order CA perturbation theory. Following such a recipe, Hordiichuk et al~\cite{Hordiichuk2023} introduced the reference system consisting of the short-range part $\,u_{\rm SR}(r)\,$ of the total LJ interaction, i.e., both the repulsion and short-range attraction energy,
\begin{eqnarray}
\label{uSR}
u_{\rm ref}(r) \equiv u_{\rm SR}(r) 
= \left\{
\begin{array}{ll}
u(r)\,, &  r \le r_{\rm m}  \\ \\
u_{\rm sr}(r)\,, &  r > r_{\rm m}
\end{array}
\right.\,.  
\end{eqnarray}
The subscript "SR"  in Eq.~(\ref{uSR}) stands for the sum of repulsive part of $u(r)$ plus short-range attractive $u_{\rm sr}(r)$ interaction energy of a target molecule and its short-range neighboring counterparts. Since repulsion in the LJ fluid is soft, to keep the short-range potential function continuous, the modeling of a short-range attraction $\,u_{\rm sr}(r)\,$ involves the two-Yukawa function in the form~\cite{Hordiichuk2023},
\begin{eqnarray}
\label{usrLJ}
u_{\rm sr}(r) =  
{\displaystyle \epsilon\frac{r_{\rm m}}{r}} 
\left [e^{\displaystyle -\alpha(r-r_{\rm m})} - 
2e^{\displaystyle -\beta(r-r_{\rm m})}\right ]\,. 
\end{eqnarray}
The two-Yukawa form provides sufficient flexibility to reproduce both the depth and curvature of the LJ potential minimum while enforcing a rapid decay beyond the first coordination shell. The decay parameters $\,\alpha\,$ and $\,\beta\,$ are determined from two conditions: (1) the distance derivative of the function $\,u_{\rm SR}(r)\,$ should be equal to zero in the potential well $\,r_{\rm m}\,$, and (2) the short-range attraction $\,u_{\rm sr}(r)\,$ should decay in such a way that at distance $\,r=r_{\rm sr}\,$, associated with the radii of the first coordination shell in a condensed LJ fluid at temperature $\,kT/\epsilon = 1\,$ and density $\,\rho\sigma_{\rm LJ}^3 = 0.85\,$ its energy will vanish (in practice equals $\,\delta k_{}T\,$, where $\,\delta = 10^{-3}\,$. 
Proceeding in this way, we obtained $\,\alpha\sigma_{\rm LJ} = 24.17\,$ and $\,\beta\sigma_{\rm LJ} = 11.63\,$. Then, the remaining perturbation attraction $\,\Delta u(r) = u(r) - u_{\rm SR}(r)\,$ reads
\begin{eqnarray}
\label{ulr}
u_{\rm pert}(r) \equiv \Delta u(r)  = \left\{
\begin{array}{llll}
0, &  r\le r_{\rm m} \\ \\
u(r) - u_{\rm sr}(r), &  r > r_{\rm m} \,.
\end{array}
\right.  
\end{eqnarray}

Figure~\ref{FigPOT} (the right panel) shows the total LJ interaction energy $\,u(r)\,$, the short-range reference interaction energy 
$\,u_{\rm SR}(r)\,$, and the perturbation energy $\,\Delta u(r)\,$, all according to their definitions by Eq.~(\ref{uLJ}) and Eqs.~(\ref{uSR})--(\ref{ulr}), respectively. It is worth emphasizing that the scheme outlined here, when applied to the LJ potential, represents a novel non-classical approach to split the pair interaction energy $\,u(r)\,$ into short-range reference $\,u_{\rm SR}(r)\,$ and  remaining perturbation $\,\Delta u(r)\,$ parts, the scheme that is in contrast to the common classical  practice~\cite{Zwanzig1954,BH1976,WCA1980}. 

\subsection{First order perturbation approximation}
In the the non-classical NCA splitting, the reference and perturbation interactions possess distinct minima. At the same time, the minimum of the short-range reference potential coincides with that of the full LJ potential. This coincidence implies that the dominant energetic and structural contributions arising from the first coordination shell are already incorporated into the reference system. The latter is an indirect indication that the properties of two systems might be close. Indeed, recently reported simulations~\cite{Hordiichuk2023} have shown that the short-range (SR) Lennard-Jones fluid  reference, in contrast to the repulsive reference, does show local structure very similar to that of the LJ fluid  not only in the dense fluid phase, but also in the gaseous phase. Moreover, its thermodynamic properties are also evidently closer to those of the full LJ fluid. 
All these observations indicate that a perturbation theory based on an SR reference should be considerably simpler because the correction term may not need to compensate the failure of the reference system to describe the thermodynamic properties of the parent fluid. These observations provide a strong a posteriori justification for truncating the perturbation expansion at first order, or at least for expecting significantly improved convergence compared to purely repulsive reference systems. The latter may serve as an indication that neglecting the second- and high-order expansion terms in Eq.~(\ref{A_PT}) is possible; or at least more justified than in the case of the repulsive reference. 

Within the first-order perturbation theory, the Helmholtz free energy then assumes the form,
\begin{eqnarray}
\frac{A}{kTN} = \frac{A_{\rm SR}}{kTN} 
+ \frac{\langle \Delta U \rangle_{\rm SR}}{kTN}\,,
\label{A_vdw}
\end{eqnarray}
where the first term $\,A_{\rm SR}\,$ in Eq.~(\ref{A_vdw})  stands for the Helmholtz free energy of the SR reference fluid. 
The first-order perturbation contribution in Eq.~(\ref{A_vdw}) can be presented in terms of the radial distribution function of the SR reference fluid $\,g_{\rm SR}(r)\,$ according to 
\begin{eqnarray} 
\label{aTrho}
a(T, \rho) = -2\pi \displaystyle\int_{0}^{\infty} g_{\rm SR}(r) \Delta u(r)r^2dr\,.
\end{eqnarray}
The function $a(T,\rho)$ represents the mean attractive contribution to the Helmholtz free energy per particle arising from the long-range tail of the interaction. Unlike the classical van der Waals energy parameter, function $a(T,\rho)$ is determined by the structure of the reference system and therefore acquires an explicit density and temperature dependence. An important novelty introduced by the SR reference is that both terms in Eq.~(\ref{A_vdw}) now depend on density and temperature. As a consequence, derivatives of the Helmholtz free energy with respect to density and temperature generate additional contributions to pressure and other thermodynamic quantities, which must be included for thermodynamic consistency. In the following section, both contributions in Eq. (\ref{A_vdw}), including the full density and temperature dependence of $a(T,\rho)$, are evaluated using computer simulations to assess the accuracy of the resulting first-order theory.

\section{Results and discussion}
In this section, we present a systematic, simulation-based assessment of the non-classical first-order perturbation theory for the Lennard-Jones fluid developed in Section 2. The analysis is carried out entirely on the basis of computer simulation data and does not rely on additional analytical approximations or adjustable parameters. The short-range Lennard-Jones reference system employed here has been characterized in detail in Ref.~\cite{Hordiichuk2023}, where its phase behavior, structure, and thermodynamic properties were shown to closely resemble those of the full Lennard-Jones fluid over a wide range of states. Building on these established properties, we focus here on evaluating how this reference performs within a first-order perturbation framework, with particular emphasis on the accuracy of thermodynamic properties derived from the Helmholtz free energy and on the role of density- and temperature-dependent contributions arising from the non-classical splitting of the interaction potential.

\subsection{Thermodynamic framework and simulation strategy}
Although perturbation theory is formulated in terms of Helmholtz free energy, the primary quantities of interest are thermodynamic properties derived from it, most notably the pressure $p(T,\rho)$ and the internal energy $E(T,\rho)$, obtained as density and temperature derivatives of $A(T,\rho)$, respectively. Within the first-order perturbation theory considered here [Eq. (\ref{A_vdw})], the evaluation of these properties for the Lennard-Jones fluid therefore requires knowledge of the corresponding quantities for the short-range reference system $p_{\mathrm{SR}}(T,\rho)$ and $E_{\mathrm{SR}}(T,\rho)$, supplemented by the appropriate density- and temperature-dependent contributions arising from the perturbation term $a(T,\rho)$.

For systems with pairwise additive interactions, all thermodynamic properties can be expressed in terms of the radial distribution function. In the present formulation, the radial distribution function of the reference system, $g_{\mathrm{SR}}(r)$, thus constitutes the central microscopic input: it fully determines both the reference contribution and the first-order perturbation term. Consequently, all quantities entering Eq. (2.10) are evaluated independently but consistently from computer simulations of the short-range reference fluid, without invoking additional analytical approximations or adjustable parameters. Namely, the pressure of the reference system,
\begin{equation} \label{pSR}
p_{\rm SR} = \frac{N}{V}\left [kT - \frac{1}{6}\frac{N}{V} \int_0^\infty \frac{\partial u_{\rm SR}(r)}{\partial r} g_{\rm SR}(r) r d\mathbf{r} \right ]\,,
\end{equation}
and the internal energy of the reference system,
\begin{equation} \label{ESR}
E_{\rm SR} = \frac 1N \left [\frac 32 kT + \frac 12\frac{N}{V} \int_0^\infty u_{\rm SR}(r) g_{\rm SR}(r) d\mathbf{r} \right ]\,.
\end{equation}

The thermodynamic domain considered in the present study corresponds to homogeneous supercritical states of the Lennard-Jones fluid. Details of the phase behavior and stability of the short-range reference system are discussed in the following subsection. Here, we emphasize that the chosen state points allow for a direct and unambiguous assessment of the intrinsic performance of the first-order non-classical perturbation theory.

As a result, the assessment presented below is free from ambiguities associated with phase coexistence or metastability of the reference fluid and provides a clean test of the perturbation theory itself. The chosen state points thus allow us to isolate the accuracy of the first-order non-classical perturbation treatment and to examine how the density- and temperature-dependent perturbation contribution $a(T,\rho)$ influences thermodynamic properties across a broad range of supercritical conditions.

Within this framework, all quantities entering the first-order perturbation expression (\ref{A_vdw}) are consistently determined from molecular dynamics (MD) simulation data for the radial distribution function $g_{\rm SR}(r)$ of the short-range reference system defined by the pair interaction $u_{\rm SR}(r)$ according to Eqs.~(\ref{uSR}) and (\ref{usrLJ}). In particular, the thermodynamic properties of the reference fluid, i.e., pressure $p_{\rm SR}(T,\rho)$, internal energy $E_{\rm SR}(T,\rho)$, and the contribution of the perturbation $a(T,\rho)$ are independently evaluated for each state point considered using Eqs.~(\ref{pSR}), (\ref{ESR}) and (\ref{aTrho}). The MD simulations were performed in a microcanonical ensemble for a system of 2000 particles with periodic boundary conditions. Production runs of $3\times 10^5$ time steps $\Delta t=0.00185$ were performed for each considered thermodynamic state. The highlighted strategy allows us to disentangle the intrinsic performance of the non-classical first-order perturbation theory from uncertainties associated with approximate theoretical inputs. In the following subsections, we first define the thermodynamic domain of interest and then analyze the behavior of the perturbation contribution and its implications for thermodynamic properties.

\subsection{Thermodynamic domain and reference-system constraints}
\begin{figure}[htb]
	\centering
	\includegraphics[width=0.5\textwidth]{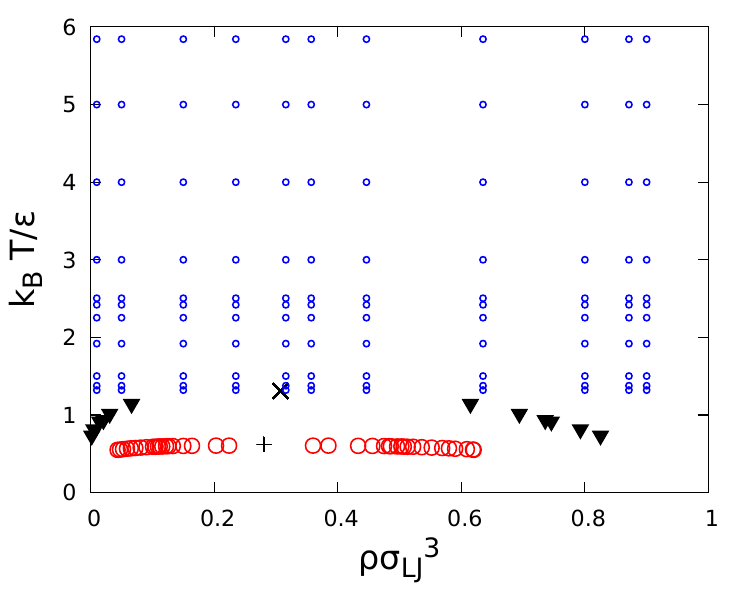}
	\caption{Vapor-liquid phase diagram of the system with short-range attraction $u_{\rm SR}(r)$ according to Eq.~(\ref{uSR}) (the empty circles) from Ref.~\cite{Hordiichuk2023} in comparison against the same for the Lennard-Jones fluid (the solid triangles) from Ref.~\cite{Alejandre1999}. The empty small circles in supercritical region indicate thermodynamic states for which MD simulations were performed.}
	\label{phase}
\end{figure}

The thermodynamic states investigated in the present study are summarized in Fig.~\ref{phase} in the temperature-density plane. The set of state points spans a broad region of the supercritical Lennard-Jones fluid, covering eleven temperatures in the range $1.321 \le T^{\star} \le 6$ and eleven densities from dilute gas-like states up to dense fluid conditions, resulting in a total of 121 simulated thermodynamic states. This domain is deliberately chosen to avoid the vicinity of phase coexistence of the parent Lennard-Jones fluid and to ensure that all simulations correspond to homogeneous equilibrium states.

Also shown in Fig.~\ref{phase} is the vapor-liquid phase diagram of the short-range Lennard-Jones reference system, reproduced from Ref.~\cite{Hordiichuk2023}, together with the corresponding coexistence region of the full Lennard-Jones fluid. As established in Ref.~\cite{Hordiichuk2023}, the critical temperature of the short-range reference fluid lies below the triple-point temperature of the Lennard-Jones fluid. This property guarantees that, throughout the entire thermodynamic domain considered here, the reference system does not undergo a vapor-liquid phase transition and remains thermodynamically stable. Consequently, the state points selected in Fig.~\ref{phase} satisfy the necessary thermodynamic constraint for the application of perturbation theory based on a state-dependent reference system.

Beyond thermodynamic stability, the applicability of first-order perturbation theory requires that the reference system reproduces the local structure of the parent fluid. As shown in Ref.~\cite{Hordiichuk2023}, this condition is satisfied for the short-range Lennard-Jones reference not only in the dense liquid regime but also in the supercritical gaseous phase. In the following subsections, we therefore focus on how these established properties of the reference system translate into the performance of the first-order non-classical perturbation theory for thermodynamic observables.

\subsection{Perturbation contribution $a(T,\rho)$: magnitude and state dependence}
We now turn to the central quantity of the non-classical first-order perturbation theory, namely the perturbation contribution $a(T,\rho)$ defined by Eq. (2.11). Within the present formulation, this function represents the mean attractive contribution arising exclusively from the long-range tail of the Lennard-Jones interaction and constitutes the sole correction to the Helmholtz free energy beyond the short-range reference system. Unlike the classical van der Waals parameter, $a(T,\rho)$ is neither constant nor purely mean-field in nature: it is determined by the structure of the reference fluid through the radial distribution function $g_{\mathrm{SR}}(r)$ and therefore depends explicitly on both density and temperature.

The behavior of the first-order perturbation term has previously been critically examined for classical WCA- and BH-based splittings by van Westen and Gross \cite{WestenGross2017}, who evaluated the perturbation contribution exactly by Monte Carlo simulations. Their analysis demonstrated that, when the entire attractive well is treated as a perturbation, the resulting $a(T,\rho)$ is large, state-dependent, and highly sensitive to density variations, leading to poor convergence of first-order perturbation theory even when higher-order terms are included. These findings clearly identify the choice of reference system -- rather than truncation of the expansion -- as the decisive factor controlling the success of perturbation theory.

In contrast, within the non-classical splitting adopted here, the perturbation interaction is confined to the weak, long-range tail of the potential, while the entire short-range part responsible for local structure is incorporated into the reference system. As a consequence, the resulting perturbation contribution $a(T,\rho)$ is expected to be significantly reduced in magnitude and to vary smoothly with the thermodynamic state. In this subsection, $a(T,\rho)$ is evaluated using computer simulation data for the short-range reference system over the entire thermodynamic domain shown in Fig. 2, allowing us to assess whether these conditions are satisfied and whether truncation of the perturbation expansion at first order is justified.

Within the first-order perturbation theory, Eq. (2.10), the contribution of the perturbation interaction $\Delta u(r)$ is fully characterized by the temperature- and density-dependent function $a(T,\rho)$. The numerical results for $a(T,\rho)$ obtained from Eq. (2.11) are shown in Figs. 3 and 4. In order to facilitate the evaluation of density and temperature derivatives required for thermodynamic consistency, these data have been represented by a smooth polynomial parameterization of the form
\begin{eqnarray}
a(T,\rho) = A_0(T)+A_1(T)\rho+A_2(T)\rho^2+A_3(T)\rho^3\,,
\label{aT}
\end{eqnarray}
%
where the temperature-dependent coefficients $A_i(T)$ are expressed as low-order polynomials in $T$, 
\begin{eqnarray}
A_i(T) = \alpha_{i0} + \alpha_{i1} T + \alpha_{i2} T^2 + \alpha_{i3} T^3 
\label{AiT}
\end{eqnarray}
with numerical values $\alpha_{ij}$ listed in Table~\ref{coefs}.

	\begin{table}[htb]
		\caption{Coefficients $\alpha_{ij}$ to describe the temperature dependence of the coefficients $A_0(T)$,   $A_1(T)$,  $A_2(T)$ and  $A_3(T)$ in Eq.(\ref{aT}) }
		\label{coefs}
		\vspace{2ex}
		\begin{center}
	\begin{tabular}{|c|cccc|}	
		\hline
  $j$\,/\,$i$&      0     & 1       & 2       & 3 \strut\\
        \hline
		0  &  4.4909 & -0.4329 &  0.1029 & -0.0082 \strut\\
		1  & -3.0671 &  2.2881 & -0.5334 &  0.0417 \strut\\
		2  &  4.0465 & -2.9527 &  0.6510 & -0.0495 \strut\\
		3  & -2.4592 &  1.3020 & -0.2546 &  0.0182 \strut\\
		\hline
	\end{tabular}
\end{center}
\end{table}

Beyond its practical utility, this parametrization reflects an important physical feature of the non-classical splitting. As illustrated in Figs. 3 and 4, the perturbation contribution $a(T,\rho)$ varies smoothly with both density and temperature and remains relatively small over the entire thermodynamic domain considered. This behavior stands in marked contrast to the results reported by van Westen and Gross \cite{WestenGross2017} for WCA- and BH-based perturbation theories, where the exact first-order perturbation term exhibits pronounced density dependence and large deviations from mean-field behavior, even when evaluated directly by Monte Carlo simulation.

\begin{figure}[htb]
	\centering
\includegraphics[width=0.75\linewidth]{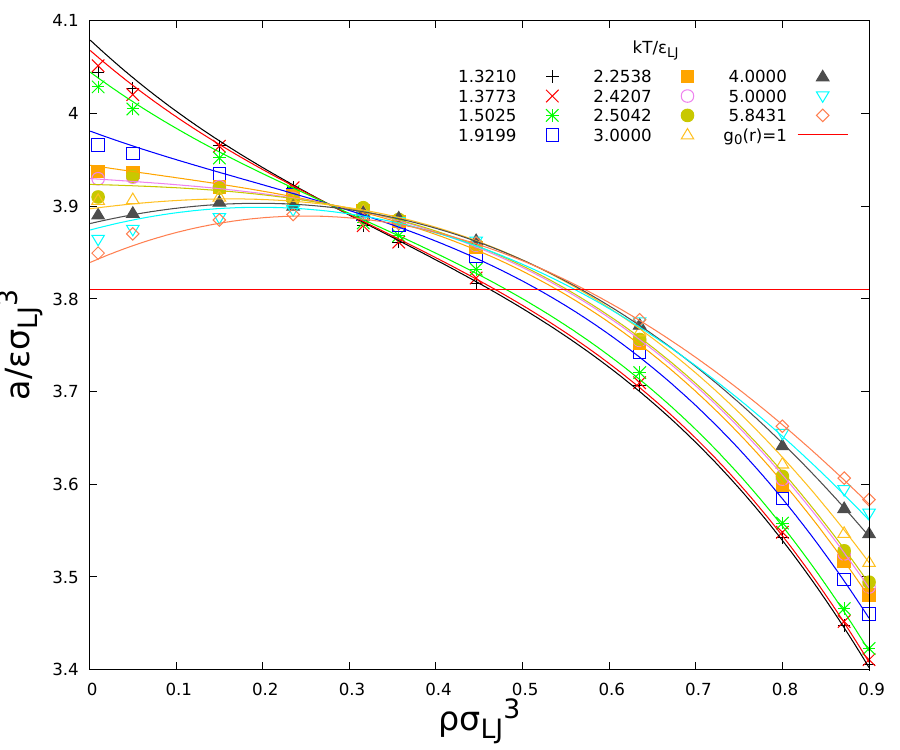}	
\caption{Dependence of the perturbation term $a(T,\rho)$ on density for the set of fixed temperatures $T^\star$ shown in figure. The results of calculations according to Eq.~(\ref{aTrho}) are marked by symbols. The thin solid curves correspond to the results of polynomial parametrization, Eq.~(\ref{aT}), while the thick solid horizontal line presents the result of mean-field approximation by assuming $\,g_{\rm SR}(r)=1\,$ in Eq.~(\ref{aTrho}).}
\label{Fig_arhoT}
\end{figure}
\begin{figure}[htb]
	\centering
	\includegraphics[width=0.75\linewidth]{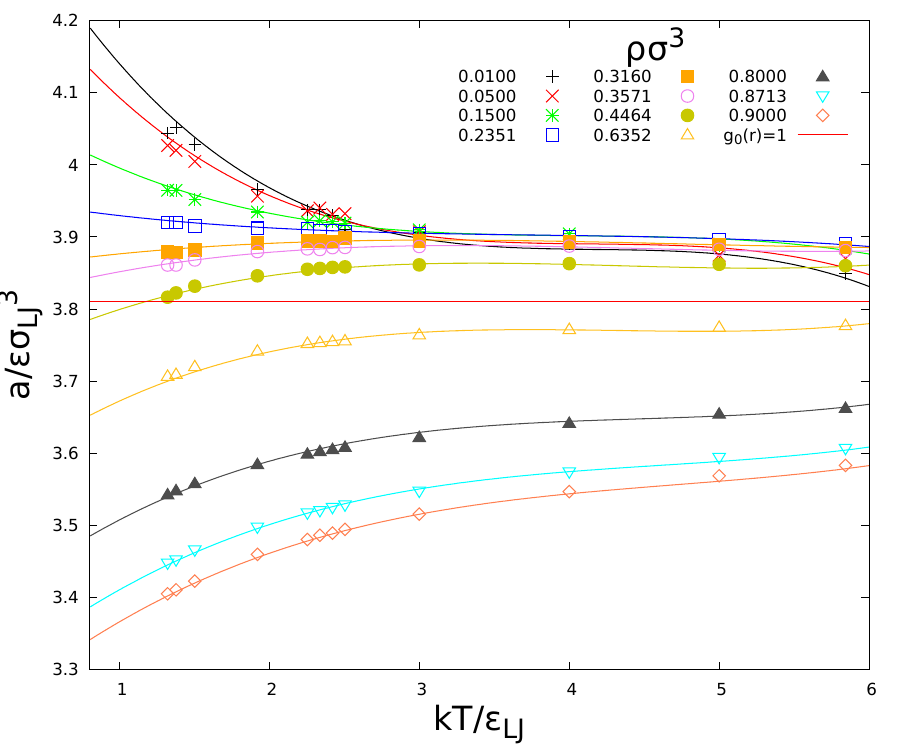}	
	\caption{ Dependence of the perturbation term $a(T,\rho)$ on temperature for the set of fixed densities $\rho^\star$ shown in figure. The results of calculations according to Eq.~(\ref{aTrho}) are marked by symbols. The thin solid curves correspond to the results of polynomial parametrization, Eq.~(\ref{aT}), while the thick solid horizontal line presents the result of mean-field approximation by assuming $\,g_{\rm SR}(r)=1\,$ in Eq.~(\ref{aTrho}).}
	\label{Fig_aTrho}
\end{figure}

The origin of this qualitative difference becomes evident when comparing the present results with the limiting case obtained by assuming $g_{\mathrm{SR}}(r)=1$ in Eq. (2.11), corresponding to a pure mean-field treatment of the perturbation. As indicated by the horizontal reference lines in Figs. 3 and 4, this approximation provides a remarkably good estimate of $a(T,\rho)$ when the perturbation interaction is defined according to the non-classical splitting, Eqs. (2.7)-(2.9). This observation implies that once the entire short-range part of the interaction responsible for local structure is absorbed into the reference system, the remaining long-range tail indeed acts as a weak, slowly varying molecular field.

In this case, the mean-field expression for the perturbation contribution,
\begin{eqnarray}
\label{avdw}
a = -2\pi\int_{\displaystyle{r_{\rm m}}}^{\displaystyle{\infty}}\Delta u(r)r^2dr 
\nonumber \\
=  -2\pi\sigma_{\rm LJ}^3\epsilon
\left [
\frac{4}{9}\left(\frac{\sigma_{\rm LJ}}{r_{\rm m}}\right)^{9} - \frac{4}{3}\left(\frac{\sigma_{\rm LJ}}{r_{\rm m}}\right)^3 - \frac{r_{\rm m}}{\sigma_{\rm LJ}} \left(
\frac{\alpha r_{\rm m} + 1}{\alpha^{2}\sigma_{\rm LJ}^2} +
2\frac{\beta r_{\rm m} + 1}{\beta^{2}\sigma_{\rm LJ}^2} \right) \right ] \,, 
\end{eqnarray}
%
provides a quantitatively accurate approximation ($a \approx  3.81\epsilon\sigma_{\rm LJ}^3$) to the exact first-order term. The relative deviation remains below $10\%$ over the entire domain. This behavior is fundamentally different from that encountered in classical perturbation approaches~\cite{WestenGross2017}, where the perturbation interaction overlaps strongly with the region of structural correlations and the mean-field approximation breaks down~\cite{MelnykJSF2010}. Therefore, the present results  demonstrate that the mean-field character of $a(T,\rho)$ is  sensitive to the choice of the reference system.

Figures 3 and 4 summarize the behavior of the perturbation contribution $a(T,\rho)$ as a function of density and temperature, respectively. Several general features are immediately apparent. First, over the entire thermodynamic domain investigated, $a(T,\rho)$ remains a smooth and slowly varying function of the state. Its magnitude changes moderately with density at a fixed temperature and decreases monotonically with increasing temperature at a fixed density, reflecting the diminishing influence of attractive interactions at higher thermal energies.

Importantly, the density dependence of $a(T,\rho)$ is relatively weak, particularly in comparison with the behavior reported for classical perturbation theories. As demonstrated by van Westen and Gross \cite{WestenGross2017}, when the entire attractive well is treated as a perturbation within WCA or Barker-Henderson splittings, the exact first-order perturbation term exhibits pronounced density sensitivity and large deviations from mean-field behavior. In contrast, the present results show that, once the short-range attractive interactions responsible for local structure are incorporated into the reference system, the remaining long-range contribution varies smoothly and does not introduce strong density-driven fluctuations.

This conclusion is further supported by the comparison with the mean-field approximation obtained by setting $g_{\mathrm{SR}}(r)=1$ in Eq. (2.11), indicated by the horizontal reference lines in Figs. 3 and 4. The close proximity of the exact simulation results to this limit demonstrates that within the non-classical splitting the perturbation term behaves to a large extent as a genuine molecular field. Deviations from the mean-field value remain small and systematic, indicating that residual structural correlations in the long-range tail play only a minor role.

Taken together, the trends observed in Figs. 3 and 4 demonstrate that, within the non-classical splitting, the perturbation contribution $a(T,\rho)$ is transformed into a weak, smooth, and near-mean-field quantity over a broad range of thermodynamic states. This behavior is not imposed by assumption, but emerges naturally once the entire short-range part of the interaction responsible for local structure is incorporated into the reference system.

Importantly, however, the small magnitude and regular state dependence of $a(T,\rho)$ alone do not guarantee the success of the first-order perturbation theory. Because the reference system is explicitly state-dependent, the thermodynamic quantities derived from the Helmholtz free energy involve derivatives of $a(T,\rho)$ with respect to density and temperature. The proper inclusion of these derivative contributions is therefore an essential element of a thermodynamically consistent formulation and must be assessed explicitly.

\subsection{Thermodynamic derivatives and consistency}
At this point, it is instructive to contrast the present results with the critical simulation-based assessment of the first-order perturbation theory reported by van Westen and Gross \cite{WestenGross2017}. In that study, the first-order perturbation contribution was evaluated exactly for classical Barker-Henderson and WCA splittings, eliminating uncertainties associated with approximate structural inputs. Despite this exact treatment, the resulting perturbation terms were found to be large, strongly state-dependent, and poorly approximated by mean-field expressions, leading to significant deviations in thermodynamic properties even when higher-order corrections were considered.

The present results demonstrate that  these shortcomings are not inherent in the first-order perturbation theory itself, but arise from the classical choice of the reference system. When the perturbation overlaps with the region controlling local structure, an exact evaluation of the first-order term does not ensure convergence. In contrast, by incorporating the entire short-range interaction into the reference system, the non-classical splitting confines the perturbation to the weak long-range tail, rendering the first-order contribution small, regular, and near-mean-field in character. This comparison highlights that the decisive factor governing the success of first-order perturbation theory is the physical definition of the reference system, rather than the formal truncation order of the expansion.

An essential consequence of the non-classical splitting adopted in the present work is the explicit dependence on the density and temperature  of the perturbation contribution $a(T,\rho)$. As a result, thermodynamic quantities derived from the Helmholtz free energy acquire additional terms originating from derivatives of $a(T,\rho)$ with respect to the thermodynamic variables. These contributions have no analog in classical perturbation theories formulated at the mean-field level and are therefore sometimes omitted in approximate formulations.

However, within the present framework, such derivative terms are not optional corrections but a direct and unavoidable consequence of the state-dependent reference system. Their inclusion is required to preserve thermodynamic consistency and to assess meaningfully the performance of the first-order perturbation theory. In the following, we examine explicitly the role of these derivative contributions for the pressure and internal energy of the Lennard-Jones fluid. 

\subsubsection*{A. Pressure} To elucidate the role of the perturbation contribution $a(T,\rho)$ and, in particular, of its density derivative, we first analyze the equation of state of the Lennard-Jones fluid. Pressure is the most sensitive thermodynamic quantity in this respect, as it directly probes density derivatives of the Helmholtz free energy and therefore provides a stringent test of thermodynamic consistency. Within the first-order perturbation approximation, Eq. (2.10), the pressure assumes the form
\begin{equation}
p = p_{\rm SR} - \rho^2\Big[a(T,\rho)+\rho\frac{\partial a(T,\rho)}{\partial \rho}\Big]\,.
\label{pLJt}
\end{equation}
where the second term explicitly reflects the state dependence of the perturbation contribution.
\begin{figure}[htb]
	\centering
	\includegraphics[width=0.75\linewidth]{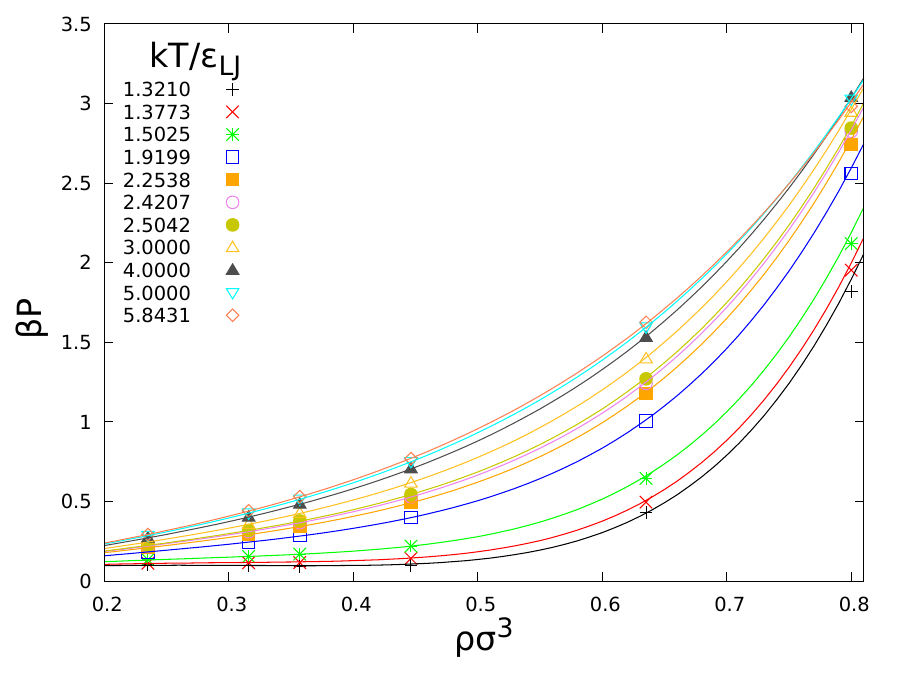}
	\caption{Dependence of the pressure  $\beta p = p/kT$ on density for the set of fixed temperatures $T^\star = kT/\epsilon_{\rm LJ}$. The solid lines are the results of of computer-based equation of state for LJ fluid \cite{KolafaNezbeda1994}, the filled symbols show results of Eq.~(\ref{pLJt}).}
	\label{pLJ}
\end{figure}

The results obtained from Eq. (3.6) are shown in Fig.\ref{pLJ} and compared with the highly accurate equation of state for the Lennard-Jones fluid due to Kolafa and Nezbeda~\cite{KolafaNezbeda1994}. This equation of state, derived from theoretical considerations combined with extensive simulation data, is widely regarded as one of the most reliable benchmarks for homogeneous Lennard-Jones fluids and has been identified as such in recent comparative assessments of the Lennard-Jones equations of state~\cite{StephanFPE2020}. As seen in Fig.\ref{pLJ}, the agreement between the present NCA-based first-order perturbation theory and the Kolafa-Nezbeda equation of state is excellent over the entire thermodynamic domain considered, from dilute supercritical states up to dense fluid conditions.
\begin{figure}[htb]
	\centering
\includegraphics[width=0.485\linewidth]{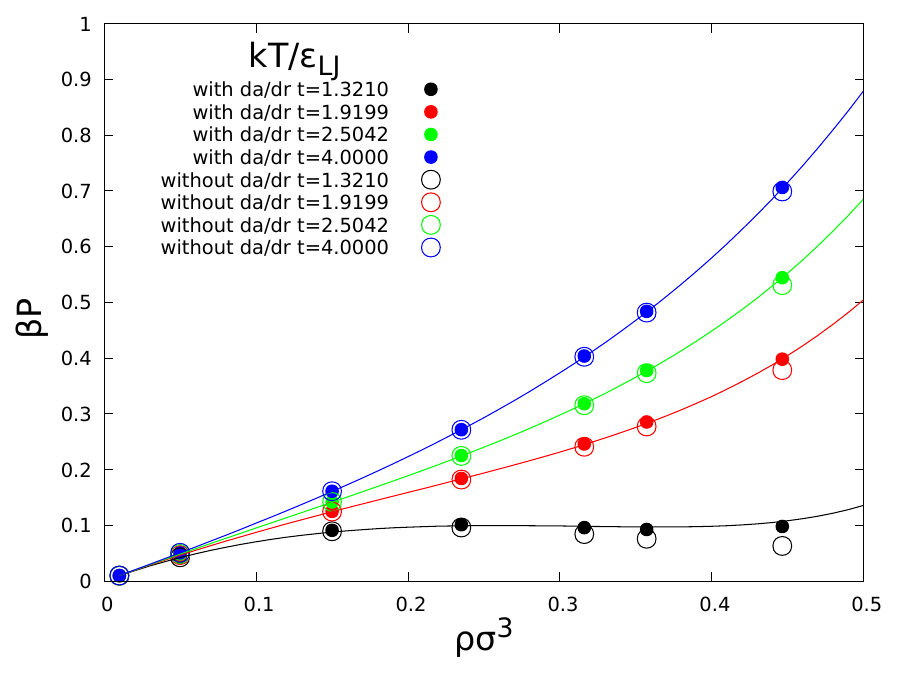}
\includegraphics[width=0.485\linewidth]{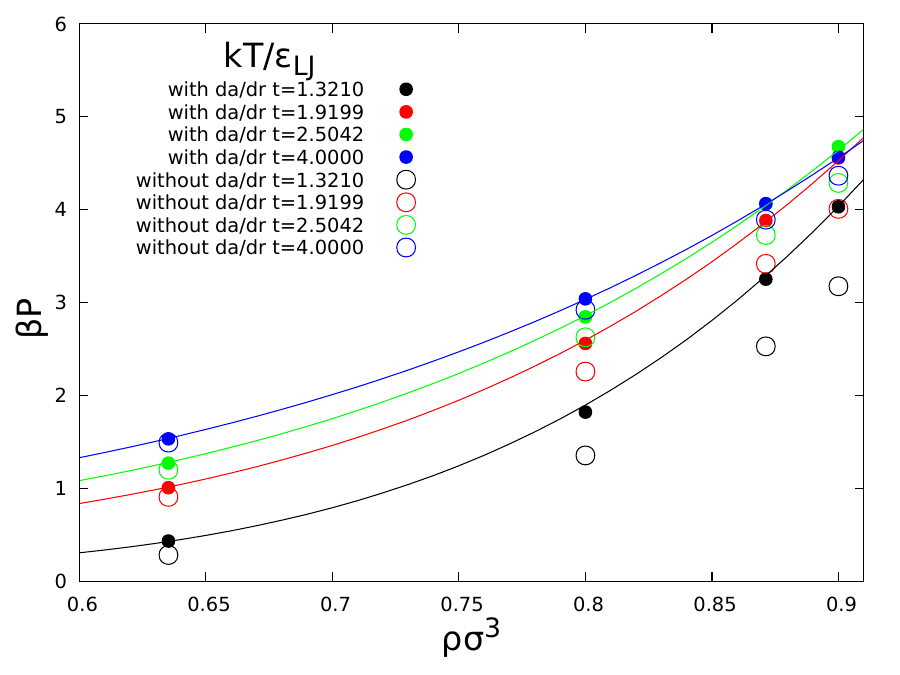}
	\caption{The same as in Fig.~\ref{pLJ} but with 
    the open symbols corresponding to negligence of the density derivative in Eq.~(\ref{pLJt}).}
	\label{pLJdiff}
\end{figure}

To isolate the role of the density-derivative term in Eq. (3.6), Fig.~\ref{pLJdiff} presents a direct comparison of pressure results obtained with and without the contribution involving $\partial a(T,\rho)/\partial\rho$. The omission of this term leads to pronounced deviations from the reference equation of state, which become particularly severe at higher densities. In contrast, inclusion of the full derivative structure restores quantitative agreement with the benchmark data. This comparison demonstrates that the density derivative of the perturbation contribution is not a small correction but an essential component of a thermodynamically consistent first-order theory when the reference system is state-dependent.

These findings underscore a central conceptual point of the present approach. The success of the non-classical first-order perturbation theory does not rely solely on the reduced magnitude of the perturbation contribution $a(T,\rho)$, but equally on the consistent treatment of its density dependence. Neglecting the derivative term would reintroduce systematic errors comparable in magnitude to those encountered in classical perturbation theories, despite the improved choice of the reference system.

\subsubsection*{B. Internal energy} 
\begin{figure}[htb]
	\centering
	\includegraphics[width=0.75\linewidth]{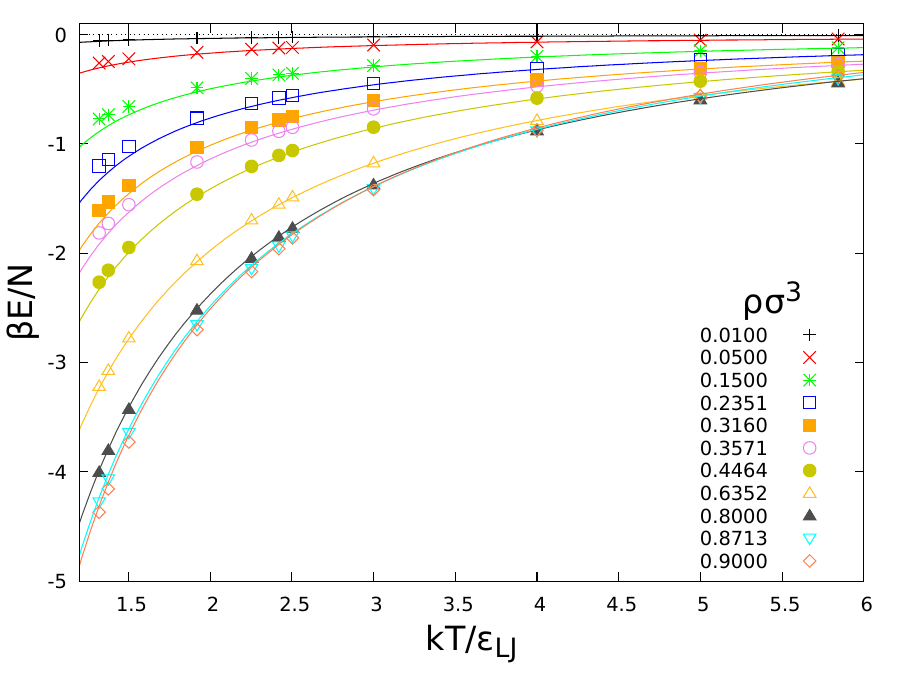}
	
	\caption{Dependence of the internal energy  $\beta E/N = E/NkT$ on temperature for the set of fixed densities $\rho^\star = \rho\sigma_{\rm LJ}^3$ shown in figure. The solid lines are the results of computer-based approach due to Kolafa and Nezbeda \cite{KolafaNezbeda1994}, the filled symbols show results of Eq.~(\ref{eLJt}).}
	\label{eLJ}
\end{figure}

The role of the temperature derivative is illustrated by the internal energy of the Lennard-Jones fluid. The equation for internal energy $E(T,\rho)$ involves the temperature derivative and can be presented in the form
\begin{equation}
\frac{E}{N} = \frac{E_{\mathrm SR}}{N}
- \rho \Big[a(T,\rho)+\frac{\rho}{kT}\frac{\partial a(T,\rho)}{\partial T^{-1}}\Big] \,.
\label{eLJt}
\end{equation}
The results of Eq.~\ref{eLJt} are shown in Fig.\ref{eLJ} in comparison against data derived due to Kolafa and Nezbeda~\cite{KolafaNezbeda1994}. Similarly to the pressure in Fig.\ref{pLJ}, the performance of Eq.~\ref{eLJt} is highly accurate in the entire domain of thermodynamic states.

In contrast to pressure, the contribution arising from the temperature derivative of $a(T,\rho)$ is negligible for the internal energy over the entire thermodynamic domain considered. 

\section{Conclusions}

In this work, we have presented a systematic assessment of a non-classical first-order perturbation theory for the Lennard-Jones fluid based on a short-range reference system that incorporates both repulsive and attractive interactions within the first coordination shell. Unlike classical perturbation approaches, where the entire attractive well is treated as a perturbation, the present formulation confines the perturbation to the weak long-range tail of the interaction potential and thereby alters the character of the first-order correction.

Using extensive computer simulation data, we have shown that the resulting perturbation contribution $a(T,\rho)$ remains small in magnitude, varies smoothly with density and temperature, and exhibits a near-mean-field character over a wide range of supercritical thermodynamic states. In contrast to classical WCA- and Barker-Henderson-based splittings, where the exact first-order perturbation term is large and strongly state-dependent, the present range-based decomposition transforms the perturbation into a weak and regular contribution.

A central outcome of this study is the explicit demonstration that the state dependence of the perturbation contribution cannot be neglected without violating thermodynamic consistency. When the density derivative of $a(T,\rho)$ is properly included, the resulting equation of state reproduces high-accuracy reference data for the Lennard-Jones fluid with excellent fidelity across a broad range of densities and temperatures. Omitting this contribution leads to systematic deviations comparable to those encountered in classical perturbation theories.

By incorporating the entire short-range interaction -- including the potential minimum -- into the reference system, the non-classical approach achieves a separation by interaction range rather than interaction strength. The improvement is not achieved by empirical parameter tuning to thermodynamic data, but follows directly from a physically motivated decomposition of the interaction range. As a result, the remaining long-range contribution genuinely acts as a weak molecular field, for which a first-order perturbation treatment becomes both accurate and thermodynamically consistent.

The present results demonstrate that, within the supercritical domain investigated, the success of first-order perturbation theory is governed not by the formal truncation order of the expansion, but by the physical content of the reference system and by the consistent treatment of its state dependence. In this sense, the classical van der Waals separation into excluded-volume and molecular-field contributions remains valid in spirit, provided that the reference system is defined on a physically meaningful microscopic basis. This framework places the van der Waals concept on a refined microscopic footing and offers a viable route toward molecular-based equations of state applicable over broad thermodynamic ranges.

\section*{Acknowledgements}
A.T. acknowledges financial
support through the MSCA4Ukraine project, funded by the European Union
(grant agreement ID: 1039539)

\end{document}